\documentclass[prb,twocolumn,showkeys,showpacs,preprintnumbers,amsmath,amssymb,floatfix]{revtex4}

\usepackage{graphicx}
\usepackage{dcolumn}
\usepackage{bm}
\usepackage{longtable}
\usepackage{ulem}

\begin{document}

\title{Scaling relation of the anomalous Hall effect in (Ga,Mn)As}

\author{M. Glunk}
\author{J. Daeubler}
\author{W. Schoch}
\author{R. Sauer}
\author{W. Limmer}
\email{wolfgang.limmer@uni-ulm.de}
\affiliation{Institut f\"ur Halbleiterphysik, Universit\"at Ulm, 89069 Ulm, Germany}


\begin{abstract}
We present magnetotransport studies performed on an extended set of (Ga,Mn)As samples
at 4.2 K with longitudinal conductivities $\sigma_{xx}$ ranging from the low- to the
high-conductivity regime. The anomalous Hall conductivity $\sigma_{xy}^{(AH)}$ is
extracted from the measured longitudinal and Hall resistivities. A transition from
$\sigma_{xy}^{(AH)}=20\;\Omega^{-1}$\,cm$^{-1}$ due to the Berry phase effect in the
high-conductivity regime to a scaling relation $\sigma_{xy}^{(AH)}\propto \sigma_{xx}^{1.6}$
for low-conductivity samples is observed. This scaling relation is consistent with a
recently developed unified theory of the anomalous Hall effect in the framework of the
Keldysh formalism. It turns out to be independent of crystallographic orientation, growth
conditions, Mn concentration, and strain, and can therefore be considered universal for
low-conductivity (Ga,Mn)As. The relation plays a crucial role when deriving values of the
hole concentration from magnetotransport measurements in low-conductivity (Ga,Mn)As. In
addition, the hole diffusion constants for the high-conductivity samples are determined
from the measured longitudinal conductivities.
\end{abstract}



\maketitle

\section{\label{introduction}Introduction}

In III-V-based diluted magnetic semiconductors such as (Ga,Mn)As, the ferromagnetic interaction
between the localized Mn spins is mediated by mobile holes.\cite{Ohno_98,Dietl_01} Spin polarization
and spin-orbit coupling of the carriers result in a large anomalous Hall effect (AHE), which has
been discussed intensively and controversially since the
1950s.\cite{Sinova_04,Nagaosa_09,Karplus_54,Smit_55_58,Luttinger_58,Berger_70}
The total Hall resistivity of ferromagnets originates from the sum of the normal Hall effect and the
AHE\cite{Berger_Bergmann_79}
\begin{equation} \label{AHE}
\rho_{yx}=\rho_{yx}^{(\rm NH)}+\rho_{yx}^{(\rm AH)}=R_0B+R_{\rm AH}\,{\rm \mu_0}M,
\end{equation}
where $R_0$ and $R_{\rm AH}$ are the ordinary and the anomalous Hall coefficients, respectively. $B$
denotes the magnetic induction and $M$ the magnetization perpendicular to the layer. Until a few years
ago, the AHE was exclusively ascribed to scattering anisotropies induced by spin-orbit
interaction,\cite{Berger_Bergmann_79} known as the extrinsic skew scattering\cite{Smit_55_58} and side
jump\cite{Berger_70} mechanisms. These mechanisms lead to contributions to $\rho_{yx}^{(\rm AH)}$ showing
a linear and quadratic dependence on the longitudinal resistivity $\rho_{xx}$, respectively.
Accordingly, Eq.~\eqref{AHE} can be rewritten as
\begin{equation}
\rho_{yx}=R_0B+c_1\rho_{xx}+c_2\rho_{xx}^2\;,
\label{r_AH_usual}
\end{equation}
where $c_1$ and $c_2$ are constants for magnetic fields high enough to saturate $M$ or, to be more
precise, to saturate the hole spin polarization. An anomalous Hall term, quadratic in $\rho_{xx}$, is also
obtained for the intrinsic (scattering independent) mechanism related to the Berry phase.\cite{Berry_84}
It was first associated with the AHE in (Ga,Mn)As by Jungwirth et al. in their seminal paper
Ref.~\onlinecite{Jungwirth_02}, where they supposed the anomalous Hall conductivity in metallic (Ga,Mn)As
to originate from strongly spin orbit-coupled holes acquiring a Berry phase during their motion on the
spin-split Fermi surfaces. Meanwhile the Berry phase mechanism is considered the dominant contribution to
the AHE in metallic (Ga,Mn)As.\cite{Jungwirth_02, Jungwirth_03, Edmonds_03, Ruzmetov_04, Chun_07, Pu_08}
Whereas most papers concerning the AHE in (Ga,Mn)As are primarily focused on the
metallic regime\cite{Jungwirth_02, Jungwirth_03, Edmonds_03, Ruzmetov_04, Chun_07, Pu_08},
only a few systematic studies concentrate on the low-conductivity regime.\cite{Shen_08, Chun_07, Allen_04}
Allen et al.\cite{Allen_04} examined the AHE in digitally doped (Ga,Mn)As structures with hopping transport
on the insulating side of the metal insulator transition (MIT). Recently, Shen et al.\cite{Shen_08} focused
on the low-conductivity regime of (Ga,Mn)As with $\rho_{xx}$ between 0.01 and 2\;$\Omega$\,cm.
They experimentally ascertained a scaling relation $\rho_{yx}^{(\rm AH)} \propto \rho_{xx}^\gamma$ with a scaling
exponent of $\gamma=0.5$. The same value for $\gamma$ was obtained in Ref.~\onlinecite{Pu_08} for a high-resistivity
(Ga,Mn)As sample showing large magnetoresistance. Neither in Ref.~\onlinecite{Pu_08} nor in Ref.~\onlinecite{Shen_08}
a theoretical explanation for this value was presented. In Ref.~\onlinecite{Chun_07}, the authors analyzed a series of
(Ga,Mn)As samples and observed a crossover of the anomalous Hall coefficient $R_{\rm AH} \propto \rho_{xx}^n$
from the low-resistivity regime dominated by the Berry phase with $n=2$ to the high-resistivity region with
$n=1.3$.\\
In this work, we experimentally find a scaling exponent of $\gamma=0.4$ in the low-conductivity
regime, which is consistent with a unified theory of the AHE, based on the Keldysh formalism,
recently developed for multiband ferromagnetic metals by Onoda et al.\cite{Onoda_06, Onoda_08}
This model predicts three regimes depending on the carrier scattering time: In the
''clean'' limit (extremely high-conductivity samples), the extrinsic skew
scattering mechanism with $\rho_{yx}^{(\rm AH)}\propto\rho_{xx}$ dominates.
This regime is beyond the conductivities achievable in the current (Ga,Mn)As growth
technology. In the intermediate regime (high-conductivity samples), the intrinsic Berry phase
mechanism with $\rho_{yx}^{(\rm AH)}\propto\rho_{xx}^2$ prevails, and in the ''dirty'' limit
(low-conductivity samples) the intrinsic contribution is strongly damped, resulting
in a scaling relation $\rho_{yx}^{(\rm AH)}\propto\rho_{xx}^{0.4}$. As shown below, the conductivities
in the (Ga,Mn)As samples under study correspond to the intermediate and dirty regime. Therefore, according
to the model of Onoda et al., Eq.~\eqref{r_AH_usual} has to be replaced by
\begin{equation}
\rho_{yx}=R_0B+c_2\rho_{xx}^2+c_3\rho_{xx}^{0.4}\;.
\label{r_AH_unusual}
\end{equation}
Whereas in nonmagnetic p-type semiconductors the hole density $p$ can be easily derived
from the ordinary Hall coefficient $R_0 = 1/ep$, obtained by magnetotransport measurements at low
magnetic fields, the determination of $p$ in (Ga,Mn)As is complicated by the dominant anomalous
contribution $\rho_{yx}^{(\rm AH)}$ in Eq.~\eqref{AHE}. To overcome this problem, $\rho_{yx}$
and $\rho_{xx}$ are usually recorded as a function of $B$ above the saturation field and
$R_0$ is derived by fitting Eq.~\eqref{AHE} to the experimental values of $\rho_{yx}$. In general,
its value is significantly affected by the choice of the relation between $\rho_{yx}^{(\rm AH)}$
and $\rho_{xx}$. In Refs.~\onlinecite{Onoda_06} and \onlinecite{Onoda_08}, the AHE is described in
terms of the anomalous Hall conductivity $\sigma_{xy}^{(\rm AH)}$ and longitudinal conductivity
$\sigma_{xx}$ instead of $\rho_{yx}^{(\rm AH)}$ and $\rho_{xx}$. Since in (Ga,Mn)As
$\rho_{yx}\ll \rho_{xx}$ has been experimentally established, the conductivities and
resistivities are related by
\begin{equation} \label{conductivity}
\sigma_{xx} = \frac{1}{\rho_{xx}}\;,\quad
\sigma_{xy}^{(\rm AH)} = \frac{\rho_{yx}^{(\rm AH)}}{\rho^2_{xx}} = \rho_{yx}^{(\rm AH)} \sigma^2_{xx}\;.
\end{equation}

\section{\label{experimental}Experimental details}

In order to investigate the scaling relation between $\sigma_{xy}^{(\rm AH)}$ and $\sigma_{xx}$
in (Ga,Mn)As, we have studied the AHE in a set of twenty samples with longitudinal conductivities
covering about one order of magnitude from 39\;$\Omega^{-1}\,$cm$^{-1}$ to 442\;$\Omega^{-1}\,$cm$^{-1}$.
The samples with thicknesses from 30 to 250\;nm and Mn concentrations between 2\% and 5\% were
grown by low-temperature molecular-beam epitaxy. Samples of different quality were obtained by variation
of the substrate temperature and/or the V/III flux ratio, as well as by using undoped semi-insulating
(001)- and (113)A GaAs wafers or partially relaxed (001) (Ga,In)As/GaAs templates. All layers showed
the typical (Ga,Mn)As reflection high energy electron diffraction (RHEED) pattern during the growth with
no visible second phase of hexagonal MnAs clusters. Further experimental details concerning the sample
growth can be found in Refs.~\onlinecite{Limmer_08,Glunk_09,Daeubler_06}.
The samples with high conductivities $\sigma_{xx}>$180\;$\Omega^{-1}\,$cm$^{-1}$ were obtained by
postgrowth annealing for 1\;h at 250\;$^\circ$C in air. Throughout this paper, samples with
$\sigma_{xx}>150\;\Omega^{-1}$\,cm$^{-1}$ at 4.2\;K are referred to as metallic.
1000\,$\mu$m$\times$300\,$\mu$m Hall bar structures were prepared from the samples by standard
photolithography and wet chemical etching. The transverse and longitudinal resistances were measured by
magnetotransport at 4.2\;K applying magnetic fields up to 14.5\;T.

\section{\label{results}Results and discussion}

\begin{figure}[h]
\centering
\includegraphics[scale=0.75]{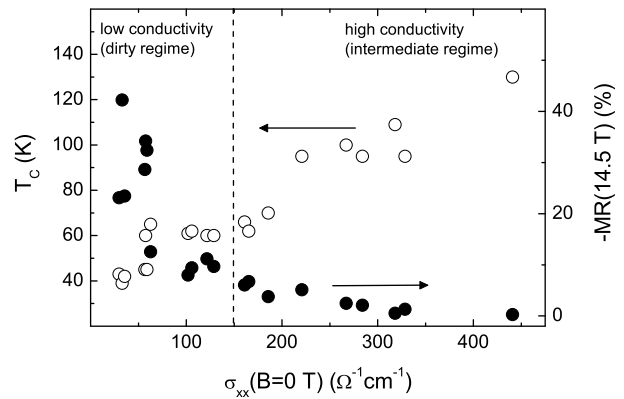}
\caption{Curie temperature $T_C$ (open circles) and absolute value of MR at
$14.5\;$T (closed circles) of the samples under study. The vertical dashed line
represents a rough boundary between the high- and low-conductivity regime.}
\label{sample_overview}
\end{figure}

\begin{figure}[h]
\centering
\includegraphics[scale=0.85]{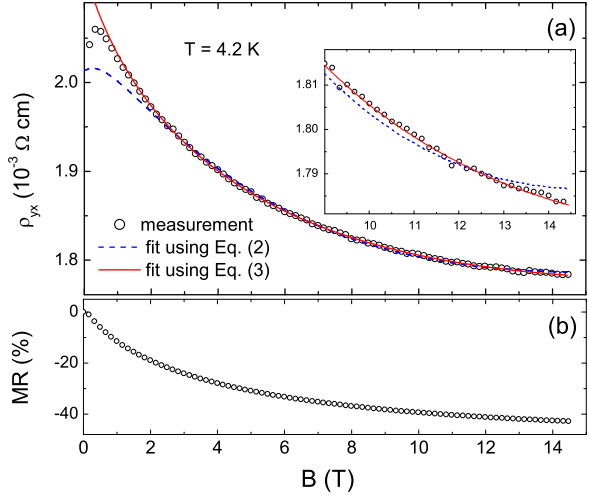}
\caption{(Color online) (a) Measured Hall resistivity of the low-conductivity sample A (open circles)
and fit curves using Eq.~\eqref{r_AH_usual} (dashed line) and Eq.~\eqref{r_AH_unusual}
(solid line). The inset shows the Hall resistivity and the fits on an enlarged scale.
(b) Magnetoresistance of sample A.}
\label{B523_113_Hall}
\end{figure}

\begin{figure}[h]
\centering
\includegraphics[scale=0.85]{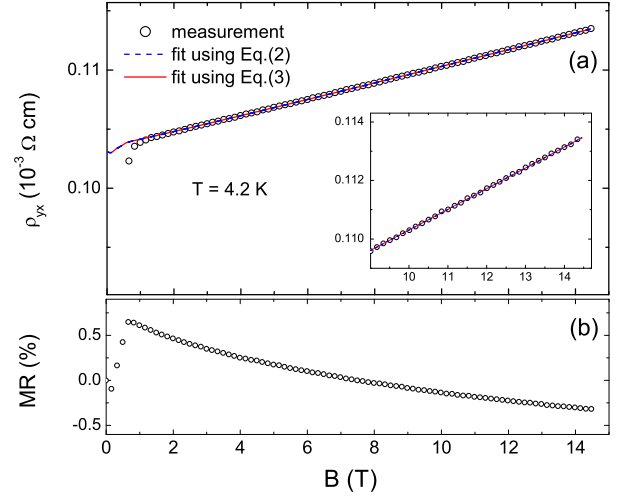}
\caption{(Color online) (a) Measured Hall resistivity of the high-conductivity sample B (open circles)
and fit curves using Eq.~\eqref{r_AH_usual} (dashed line) and Eq.~\eqref{r_AH_unusual}
(solid line). In this case the fits according to Eqs.~\eqref{r_AH_usual} and
\eqref{r_AH_unusual} fall together. The inset shows the Hall resistivity and the
fits on an enlarged scale. (b) Magnetoresistance of sample B.}
\label{B464_2_Hall}
\end{figure}

\begin{figure}[h]
\centering
\includegraphics[scale=0.85]{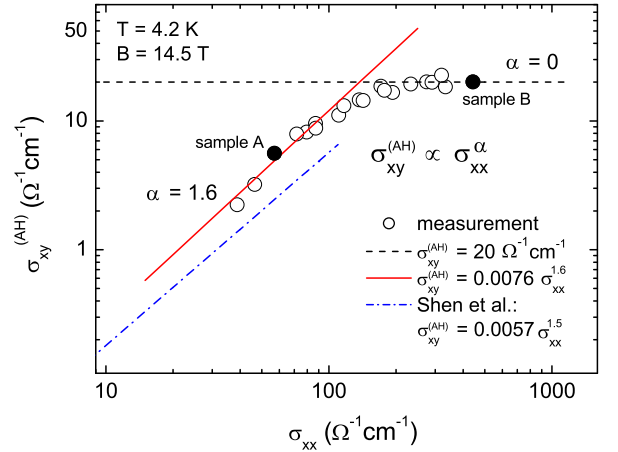}
\caption{(Color online) Anomalous Hall conductivity $\sigma_{xy}^{(\rm AH)}$ as a function of the longitudinal
conductivity $\sigma_{xx}$ at 14.5$\;$T and 4.2$\;$K. The solid line represents a universal scaling
$\sigma_{xy}^{(\rm AH)}\propto\sigma_{xx}^{1.6}$. The dashed dotted line indicates the scaling given in
Ref. \onlinecite{Shen_08}. The plateau value of $\sigma_{xy}^{(\rm AH)}= 20\;\Omega^{-1}\,$cm$^{-1}$
is attributed to the Berry phase effect.}
\label{sxy_AHE_sxx}
\end{figure}

\begin{figure}[h]
\centering
\includegraphics[scale=0.8]{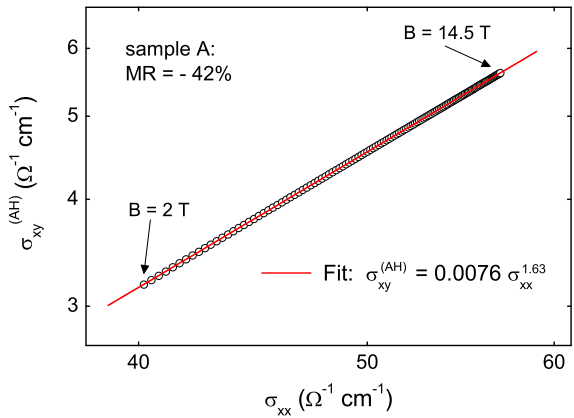}
\caption{(Color online) log-log plot of the anomalous Hall conductivity $\sigma_{xy}^{(\rm AH)}$ of the low-conductivity sample A
as a function of the magnetoconductivity $\sigma_{xx}(B)$ at 4.2$\;$K. The solid line represents a fit yielding
$\sigma_{xy}^{(\rm AH)}=0.0076\,\sigma_{xx}^{1.63}$.}
\label{sxy_sxx(B)}
\end{figure}

In Fig.~\ref{sample_overview}, the Curie temperature $T_C$, estimated from the maximum of the temperature
dependent longitudinal resistivity $\rho_{xx}$,\cite{Matsukura_98} and the magnetoresistance (MR)
$\Delta\rho_{xx}(B)/\rho_{xx}(0\,$T) at $B$=14.5\;T of the samples under study are plotted as a function of the
zero-field longitudinal conductivity $\sigma_{xx}(B=0\;T)$.
While $T_C$ increases with increasing $\sigma_{xx}$, the absolute value of the MR decreases and drops below
7\% in the metallic samples.\\
Figures~\ref{B523_113_Hall} and \ref{B464_2_Hall} show the measured Hall resistivity $\rho_{yx}$ and the MR of two
representative samples A and B, covering the cases of low ($\sigma_{xx}\approx 57\;\Omega^{-1}$\;cm$^{-1}$) and high
longitudinal conductivity ($\sigma_{xx}\approx442\;\Omega^{-1}$\,cm$^{-1}$) at 14.5\;T, respectively.
The increase of $\rho_{yx}$ and $\rho_{xx}$ below $\sim$1\;T stems from the anisotropic magnetoresistance.\cite{Limmer_08}
The Hall resistivity of sample A shown in Fig.~\ref{B523_113_Hall}(a) is dominated by the large negative MR, which can be
attributed to the suppression of weak localization effects in the presence of a magnetic field $B$.\cite{Matsukura_04}
In order to evaluate the applicability of Eqs.~\eqref{r_AH_usual} and \eqref{r_AH_unusual}, both equations were
used to fit the experimental Hall data between 2 and 14.5\;T. The inset of Fig.~\ref{B523_113_Hall}(a) clearly shows
that the measured curve is better reproduced using Eq.~\eqref{r_AH_unusual} (solid line), with $c_3\rho_{xx}^{0.4}$
being the dominant term, than using Eq.~\eqref{r_AH_usual} (dashed line). Furthermore, the dashed curve underestimates
the measurement below 2.5\;T, unphysically implying that $M$ increases with decreasing $B$. The fit using
Eq.~\eqref{r_AH_unusual} yields a hole concentration of $p=1.8\times10^{20}\;$cm$^{-3}$, in contrast to the less reasonable
low value of $p=7\times10^{19}\;$cm$^{-3}$ obtained from Eq.~\eqref{r_AH_usual}. Note that the same values are
obtained if the fit is started at magnetic fields higher than 2 T.
Comparison between Fig.~\ref{B523_113_Hall}(b) and \ref{B464_2_Hall}(b) shows that the MR for $B>2$\;T in the metallic
sample B is about two orders of magnitude smaller than in sample A, resulting in an extremely weak dependence of
$\rho_{xx}$ on $B$. Therefore, the AHE makes only a marginal contribution to the slope of $\rho_{yx}$ and, as a
consequence, Eqs.~\eqref{r_AH_usual} and \eqref{r_AH_unusual} yield the same fit curves as demonstrated in the inset of
Fig.~\ref{B464_2_Hall}(a). In both cases, a hole concentration of $p=8.5\times10^{20}\;$cm$^{-3}$ is derived. Generally, we
find that in metallic samples with a sufficiently small MR, fits to $\rho_{yx}$ yield the same values for $\rho_{yx}^{(\rm AH)}$
and $p$ irrespective of the exponents used for $\rho_{xx}$ making it impossible to separate the different contributions to
the AHE. Therefore, Eq.~\eqref{r_AH_unusual} can be used to model the $\rho_{yx}$ curves for the whole set of (Ga,Mn)As samples
under study.\\
Figure~\ref{sxy_AHE_sxx} illustrates the main finding of this paper. It shows the anomalous Hall conductivity
$\sigma_{xy}^{(\rm AH)}$ as a function of $\sigma_{xx}$ at 14.5$\;$T, determined from the measured resistivities using
Eqs.~\eqref{r_AH_unusual} and \eqref{conductivity}.
Irrespective of the growth conditions, Mn concentration, and substrate orientation, the data for the low-conductivity
samples ($\sigma_{xx}< 100\;\Omega^{-1}\,$cm$^{-1}$) follow the same universal scaling relation
\begin{equation}
\sigma_{xy}^{(\rm AH)}=s \, \sigma_{xx}^\alpha,
\label{scaling_relation}
\end{equation}
with $s=0.0076\;\rm \Omega^{0.6}\,cm^{0.6}$ and $\alpha=2-\gamma=1.6$. This value of $\alpha$ is in agreement with the
theoretical predictions made in Refs.~\onlinecite{Onoda_06} and \onlinecite{Onoda_08} for multiband ferromagnetic materials
in the ''dirty'' limit. As mentioned in Sec.~\ref{introduction}, the data presented by Shen et al.\cite{Shen_08} for
low-conductivity (Ga,Mn)As follows a very similar scaling relation $\sigma_{xy}^{(\rm AH)} \propto \sigma_{xx}^{1.5}$
(dashed-dotted line), which they claim to be incompatible with existing scattering theories. If the MR of a sample is
large, the scaling relation in Eq.~\eqref{scaling_relation} can also be tested by plotting $\sigma_{xy}^{(\rm AH)}(B)$ as a
function of the corresponding magnetic-field dependent longitudinal conductivity $\sigma_{xx}(B)$. Figure~\ref{sxy_sxx(B)}
shows the appropriate log-log plot for the low-conductivity sample A which exhibits the largest MR(14.5\,T)=-42\% of the
samples under study.
We obtain $\alpha=1.63$ ($\gamma=0.37$) in good agreement with the aforementioned theory and the value $\gamma=0.5$
derived by Pu et al.\cite{Pu_08} from a low-conductivity (Ga,Mn)As sample. Interestingly, even for the insulating (Ga,Mn)As
samples with hopping transport as studied in Ref.~\onlinecite{Allen_04} the scaling exponent $\gamma$ turned out to be close
to 0.4. Due to the fact that the scaling exponent $\alpha=1.6$, or equivalently $\gamma=0.4$, has also been found in numerous
other low-conductivity ferromagnets, independent of the details of the underlying transport process
(see e.g. Refs.~\onlinecite{Onoda_06, Onoda_08, Nagaosa_09}, and references therein, and
Refs.~\onlinecite{Miyasato_07, Ueno_07, Fernandez_08, Sangio_09}), we are led to the conclusion that it has to be considered universal
and applies to low-conductivity (Ga,Mn)As irrespective of the specific material parameters. For samples with longitudinal
conductivities between $\sim$100 and $\sim$200\;$\Omega^{-1}$\,cm$^{-1}$, $\alpha$ gradually decreases to $\alpha=0$ with
increasing $\sigma_{xx}$ in qualitative agreement with the theory presented in Refs.~\onlinecite{Onoda_06} and
\onlinecite{Onoda_08}. For the high-conductivity samples with $\sigma_{xx}>200$\;$\Omega^{-1}$\,cm$^{-1}$, $\sigma_{xy}$
becomes independent of $\sigma_{xx}$ amounting to $\sigma_{xy}^{(\rm AH)}=20\;\Omega^{-1}$\,cm$^{-1}$.
This value is in quantitative agreement with theoretical data in Ref.~\onlinecite{Jungwirth_03} calculated
within the framework of the Berry phase theory of the AHE for a disordered sample with $p=6.0\times10^{20}\;$cm$^{-3}$.
Since the mean value of $p$ derived for the metallic samples under study is also close to $6.0\times10^{20}\;$cm$^{-3}$
(see below), our result for the high-conductivity regime corroborates the intrinsic origin of the AHE in metallic (Ga,Mn)As.
As demonstrated in Fig.~\ref{sxy_AHE_sxx}, a significant linear contribution $\sigma_{xy}^{(\rm AH)} \propto \sigma_{xx}$ due
to skew scattering can be ruled out for the investigated range of $\sigma_{xx}$. This is in agreement with the calculations
of Nagaosa et al.\cite{Nagaosa_09} giving an estimate $\sigma_{xx}\gg 1000\;\Omega^{-1}$\,cm$^{-1}$ for the onset of dominant
skew scattering mechanism.

\begin{figure}[h]
\centering
\includegraphics[scale=0.75]{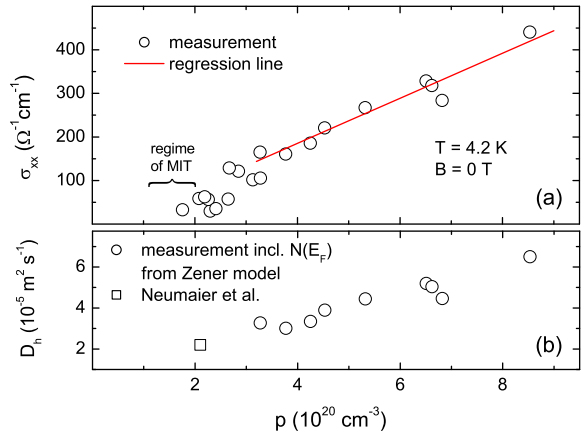}
\caption{(Color online) (a) Longitudinal conductivity $\sigma_{xx}$ at 0$\;$T as a function of the hole concentration $p$.
The regression line indicates the proportionality $D_h N_h(E_F)\propto p$ in the metallic regime.
(b) Open circles represent the diffusion constant of the holes derived from Eq.\eqref{Einstein_sigma}.
The value extracted by Neumaier et al.\cite{Neumaier_09} is indicated by an open square.}
\label{mobility}
\end{figure}

\begin{figure}[h]
\centering
\includegraphics[scale=0.75]{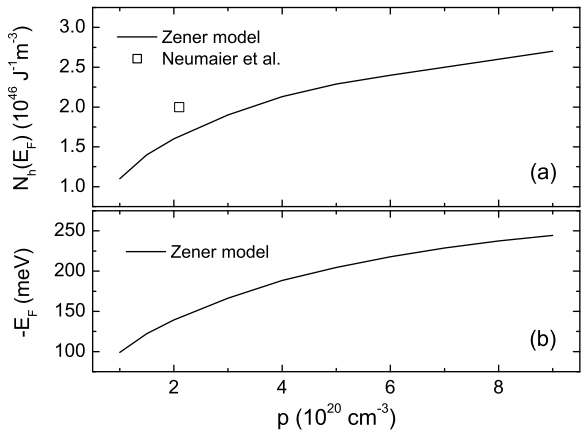}
\caption{(a) DOS $N_h(E_F)$ at the Fermi energy in (Ga,Mn)As calculated by means of the Zener model of
Dietl et al.\cite{Dietl_01} The value experimentally derived from Neumaier et al.\cite{Neumaier_09} is
indicated by an open square. (b) Absolute value of the Fermi energy $E_F$ calculated as a function of
the hole density $p$.}
\label{DOS}
\end{figure}

In the remainder of this work we focus on the longitudinal charge transport. In Fig.~\ref{mobility}(a), $\sigma_{xx}$
measured at $B=0$\;T is plotted as a function of $p$.
From the data we expect the MIT to occur between $p\approx1\times10^{20}\;$cm$^{-3}$ and $p\approx2\times10^{20}\;$cm$^{-3}$.
The simple relation $\sigma_{xx}=e p \mu_h$ yields a hole mobility of $\mu_h=3\;$cm$^2$V$^{-1}$s$^{-1}$, inferred from the slope of
the regression line (straight line) for the metallic samples, in satisfactory agreement with the values given in
Refs.~\onlinecite{Limmer_02} and~\onlinecite{Limmer_05}. In fact, the adequate description of the hole transport in metallic (Ga,Mn)As
at 4.2\;K, neclecting small quantum corrections, is given by the generalized Einstein relation\cite{Kubo_57,Imry_97}
\begin{equation}
\sigma_{xx}=e^2D_h N_h(E_F)\;,
\label{Einstein_sigma}
\end{equation}
where $D_h$ denotes the hole diffusion constant and $N_h(E_F)$ the density of states (DOS) at the Fermi energy $E_F$.
Thus, the experimentally observed linear increase of $\sigma_{xx}$ with increasing $p$ in the metallic regime indicates a
proportionality $D_h N_h(E_F)\propto p$. Only few papers\cite{Neumaier_08, Neumaier_09} have been published so far concerning
the hole diffusion constants in (Ga,Mn)As in one- and two-dimensional systems, and only little is known about the values of
$D_h$ in bulk (Ga,Mn)As. Using Eq.~\eqref{Einstein_sigma} and calculating $N_h(E_F)$ in the framework of a $6\times6$ $k \cdot p$
Zener model,\cite{Dietl_01} we have determined the hole diffusion constant $D_h$ of the metallic samples as a function of $p$.
Figure~\ref{mobility}(b) shows that $D_h$ varies between $\sim 3.0\times 10^{-5}$\;m$^2$s$^{-1}$ and
$\sim 6.5\times 10^{-5}$\;m$^2$s$^{-1}$ for the metallic samples. The extrapolation of these values to lower hole concentrations
agrees with the one found by Neumaier et al.\cite{Neumaier_09} The values $N_h(E_F)$ used in Eq.\eqref{Einstein_sigma} are shown
in Fig.~\ref{DOS}(a).
The absolute value of the Fermi energy, extracted from the same calculations, is displayed in
Fig.~\ref{DOS}(b). The exchange splitting parameter $B_G$, primarily influencing the values of $N_h(E_F)$
and $E_F$, was taken proportional to $p$ as suggested by the study in Ref. \onlinecite{Glunk_09}.
The calculations yield an increase of $|E_F|$ and $N_h(E_F)$ from $\sim$180 meV to $\sim$240 meV and from
$\sim 2.0\times 10^{46}$\;J$^{-1}$m$^{-3}$ to $\sim 2.6\times 10^{46}$\;J$^{-1}$m$^{-3}$ with increasing
$p> 3.5\times10^{20}\;$cm$^{-3}$, respectively.

\section{Summary}
The AHE has been studied in an extended set of (Ga,Mn)As samples with longitudinal conductivities $\sigma_{xx}$ covering about
one order of magnitude. For the samples in the low-conductivity regime, a universal scaling relation
$\sigma_{xy}^{(\rm AH)}\propto \sigma_{xx}^{1.6}$ or equivalently, $\rho_{yx}^{(\rm AH)}\propto \rho_{xx}^{0.4}$ is found, which
can be modeled within a fully quantum-mechanical transport theory for multiband systems recently developed in
Refs.~\onlinecite{Onoda_06} and \onlinecite{Onoda_08}. In the metallic regime, the scaling changes to a scattering independent
anomalous Hall conductivity $\sigma_{xy}^{(\rm AH)}=20\;\Omega^{-1}$\,cm$^{-1}$
($\rho_{yx}^{(\rm AH)}= 20\;\Omega^{-1}$\,cm$^{-1}\times \rho_{xx}^2$) due to the Berry phase effect. The skew scattering mechanism
can be ruled out to play a significant role for the range of $\sigma_{xx}$ under study. Consequently, Eq.~\eqref{r_AH_unusual}
instead of Eq.~\eqref{r_AH_usual}, commonly cited in the literature,  has to be used for the determination of the hole concentration
in high- {\it and} low-conductivity (Ga,Mn)As from high-field magnetotransport data.\\
In addition, hole diffusion constants between $3.0\times 10^{-5}$\;m$^2$s$^{-1}$ and $6.5\times 10^{-5}$\;m$^2$s$^{-1}$ for the metallic
samples are obtained using the measured values of $\sigma_{xx}$ and calculating the corresponding density of states $N_h(E_F)$ at
the Fermi energy within the Zener model.\\[2.0cm]
\begin{acknowledgments}
This work was supported by the Deutsche Forschungsgemeinschaft under Contract No. Li 988/4.
\end{acknowledgments}

\end{document}